\documentclass[reprint,twocolumn,prl,aps,showpacs,floatfix,amsmath, superscriptaddress]{revtex4-1}

\pdfoutput=1
\usepackage[english]{babel}
\usepackage{graphicx}
\usepackage{epsfig}
\usepackage{color}
\usepackage{amssymb}
\usepackage{amsmath}
\usepackage{array}
\usepackage[loose]{units}
\usepackage{xr-hyper}
\usepackage{hyperref}
\hypersetup{breaklinks=true}
\usepackage{braket}%
\usepackage[caption=false]{subfig}
\usepackage{multirow}
\usepackage{cleveref}
\usepackage{adjustbox}
\usepackage{yfonts}
\usepackage{letltxmacro}
\usepackage{calligra}

\LetLtxMacro{\ORIGselectlanguage}{\selectlanguage}
\makeatletter
\DeclareRobustCommand{\selectlanguage}[1]{%
    \@ifundefined{alias@\string#1}
      {\ORIGselectlanguage{#1}}
      {\begingroup\edef\x{\endgroup
         \noexpand\ORIGselectlanguage{\@nameuse{alias@#1}}}\x}%
}
\newcommand{\definelanguagealias}[2]{%
  \@namedef{alias@#1}{#2}%
}
\makeatother

\newcommand{\UO}{Department of Physics, 1274 University of Oregon, Eugene, Oregon, 97403}
\newcommand{\NCEM}{National Center for Electron Microscopy, Lawrence Berkeley National Laboratory, Berkeley, California}
\newcommand{\Gottingen}{Georg-August-Universit{\"a}t G{\"o}ttingen, IV. Physicalisches Insitut, Friedrich-Hund-Platz 1, 37077 G{\"o}ttingen, Germany}

\newcommand{\angstrom}{\mbox{\normalfont\AA}}

\begin{document}    

\author{Fehmi S. Yasin}
\affiliation{\UO}
\author{Tyler R. Harvey} 
\affiliation{\UO}
\affiliation{\Gottingen}
\author{Jordan J. Chess}
\affiliation{\UO}
\author{Jordan S. Pierce}
\affiliation{\UO}
\author{Colin Ophus}
\affiliation{\NCEM}
\author{Peter Ercius}
\affiliation{\NCEM}
\author{Benjamin J. McMorran}
\affiliation{\UO}
\email{Correspondence should be addressed to: mcmorran@uoregon.edu}

\title{Probing Light Atoms at Sub-nanometer Resolution: Realization of Scanning Transmission Electron Microscope Holography}

\date{\today}

                      
\begin{abstract}

    Atomic resolution imaging in transmission electron microscopy (TEM) and scanning TEM (STEM) of light elements in electron-transparent materials has long been a challenge. Biomolecular materials, for example, are rapidly altered when illuminated with electrons. These issues have driven the development of TEM and STEM techniques that enable the structural analysis of electron beam-sensitive and weakly scattering nano-materials. Here, we demonstrate such a technique, STEM holography, capable of absolute phase and amplitude object wave measurement with respect to a vacuum reference wave. We use an amplitude-dividing nanofabricated grating to prepare multiple spatially separated electron diffraction probe beams focused at the sample plane, such that one beam transmits through the specimen while the others pass through vacuum. We raster-scan the diffracted probes over the region of interest. We configure the post-specimen imaging system of the microscope to diffraction mode, overlapping the probes to form an interference pattern at the detector. Using a fast-readout, direct electron detector, we record and analyze the interference fringes at each position in a 2D raster scan to reconstruct the complex transfer function of the specimen, $t(\textbf{x})$. We apply this technique to image a standard target specimen consisting of gold nanoparticles on a thin amorphous carbon substrate, and demonstrate $\unit[2.4]{\angstrom}$ resolution phase images. We find that STEM holography  offers higher phase-contrast of the amorphous material while maintaining Au atomic lattice resolution when compared with high angle annular dark field STEM.

\end{abstract}

\maketitle

\section{Introduction}

    Phase contrast for low-atomic-number, beam-sensitive materials has long been pursued in electron microscopy, seeing the advent of multiple transmission electron microscopy (TEM) and scanning TEM (STEM) techniques over the past 60+ years, including electron holography or interferometry using both wavefront-dividing beamsplitters \cite{gabor_1948, haine_and_mulvey_1952, midgley_2009, hytch_2008, Cooper_2014, lichte_1991, tanigaki_2012} and amplitude-dividing beamsplitters \cite{matteucci_amplitude_1981, ru_electron_1994, zhou_principle_2001, marton_1952, marton_1953, marton_1954, yasin_2018}, ptychography \cite{rodenburg_experimental_1993, pennycook_efficient_2015, yang_simultaneous_2016, yang_enhanced_2016, pelz_low-dose_2017}, cryo-electron microscopy \cite{li_electron_2013, elbaum_cryo-scanning_2016, pelz_low-dose_2017}, matched illumination and detector interferometry \cite{ophus_efficient_2016, yang_enhanced_2016}, differential phase contrast \cite{mccartney_differential_1996, denneulin_differential_2016}, and more. These techniques have benefited from the development of technologies such as fast readout detectors and aberration correctors that have driven imaging resolution of STEM below 0.41 \angstrom \cite{morishita_2018} and TEM below 0.43 \angstrom \cite{akashi_aberration_2015}. 
    
    Several decades ago, an interferometric technique called STEM holography (STEMH) was initially developed as a phase contrast electron imaging technique \cite{leuthner_1989, cowley_STEMH_1990, takahashi_1994, cowley_STEMH_2003}. These arrangements used a charged biprism wire to split an electron beam into two probes focused at the specimen. With one beam transmitted through the specimen, the interference between the two was recorded. Due to the slow throughput and limited geometries of detectors at the time, STEMH was never widely implemented. The recent advent of fast-readout direct electron detectors enables STEMH as a practical imaging technique. Additionally, advances in FIB fabrication technologies allowed us to expand on this technique with the addition of a static, nanofabricated, amplitude-dividing diffraction grating for use as probe-forming aperture and beam splitter in a multiple-path-separated interferometer \cite{yasin_2018}. In this article, we provide such a demonstration.

    Amplitude-dividing beamsplitters in the form of nanofabricated electron diffraction gratings have been developed by multiple groups. In contrast to wavefront-dividing beamsplitters such as electrostatic biprisms, these diffraction gratings lower the coherence width requirements of the beam, while also allowing for careful shaping of the electron wave front’s phase and amplitude structure \cite{verbeeck_2010, mcmorran_2011, grillo_2014, harvey_2014, shiloh_2014}. They form symmetric profile probes at the specimen plane (grating's diffraction plane) that are absent of any unwanted edge-diffraction artifacts, and have one passive working part equal in size and shape to conventional apertures, making them easily installable into commercial electron microscopes. Additionally, although many diffraction order probes are generated from the grating, the diffraction efficiency of the grating can be tuned to decrease the intensity in the higher orders \cite{harvey_2014, yasin_2018}. 

    Another technological advance that enables STEMH is the advent of fast-readout direct electron detectors. These detectors are capable of acquiring thousands of images in seconds and are sensitive to individual electrons. Such a fast readout is necessary for any high resolution 4D-STEM imaging technique. In addition to a fast readout of $\unit[10^{2}]{\mathrm{fps}}$, the high detective quantum efficiency of such detectors should allow for a decrease in the electron dose seen by the specimen by at least two orders of magnitude \cite{ruskin_2013}. STEMH combines the aforementioned direct electron detector, amplitude-division diffraction gratings, interference fringe phase reconstruction, and aberration correction to provide quantitative phase contrast, including the dc-component with respect to vacuum.    

    In this article, we provide the theoretical framework for a three-beam, path-separated electron interferometer with a phase imparted onto one or more paths. We then provide two proof-of-principle STEMH images of Au on C, with high-angle annular dark field (HAADF) images for comparison. In HAADF STEM, the beam current is focused to a sub-nanometer width and is scanned across a field of view, dwelling at each location until a sufficient number of high-angle scattering events have illuminated an annulus detector, forming contrast. In STEMH, we extract the phase contrast in these images from the data using the aformentioned model, and find that the phase structure calculated in the amorphous carbon region is consistent with the thick-bonding theoretical model proposed by Ricolleau et al. \cite{ricolleau_2013}.

\begin{figure}[h]

\centering
  \includegraphics[width=0.99\columnwidth]{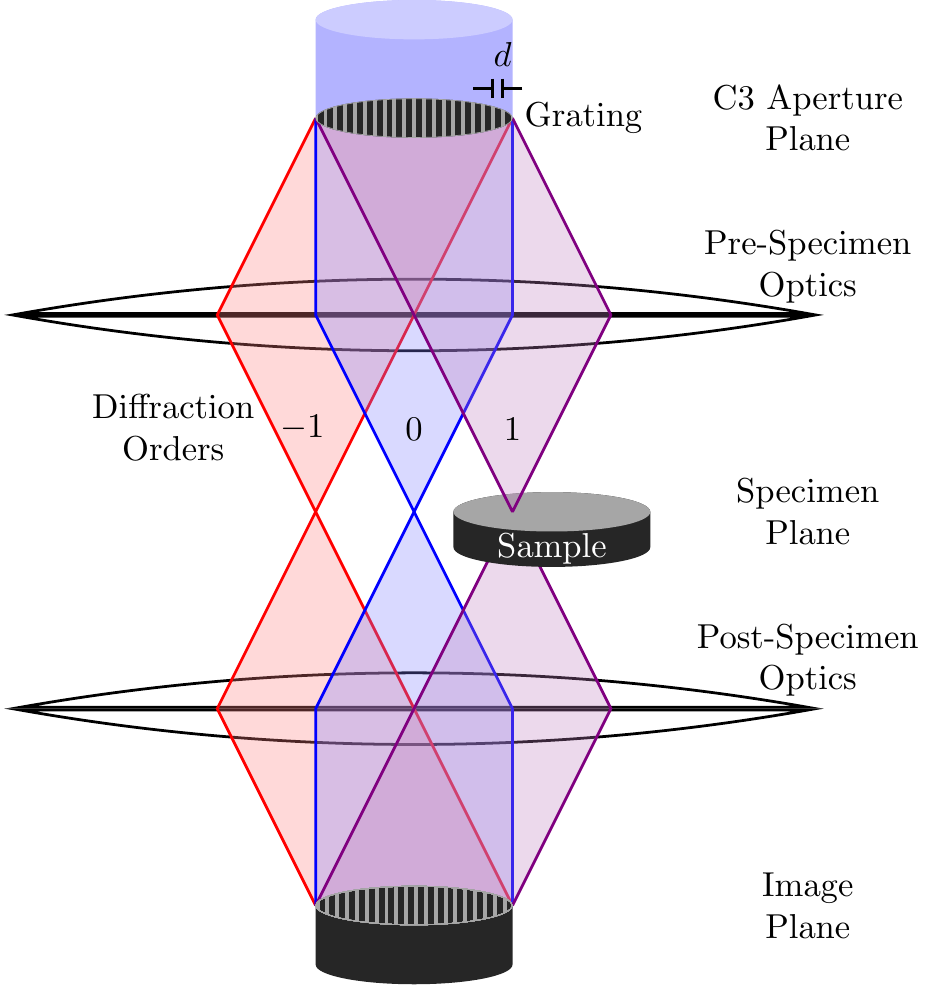} 
  \caption{
    STEM holography electron optical setup.
  \label{fig:setup} }

\end{figure}

\section{Experimental Setup}

    As illustrated in Fig. \ref{fig:setup}, the input plane wave electron beam travels down the microscope column to the probe-forming aperture, where a diffraction grating coherently splits the electron beam into multiple diffraction orders that are sharply peaked at the specimen plane, with tens of nanometers spatial separation. The specimen is positioned such that all three diffraction probes, which we'll call $probe_{+1}$, $probe_{0}$, and $probe_{-1}$ in the text, initially pass through vacuum. These probes are then rastered across the field of view along the same line as the diffraction pattern's orientation using the scanning (deflection) coils in the microscope. $probe_{+1}$ interacts with the specimen while $probe_{0}$ and $probe_{-1}$ pass through vacuum, acting as reference beams in three parallel interferometers. An interference pattern is focused onto the detector, and the fringes shift as the phase imparted onto the interacting probe varies.

\begin{figure}[hb]
    \centering
    \begin{tabular}{cc}

    \subfloat[]{\label{subfig:im1}
      \includegraphics[width=0.5\columnwidth]{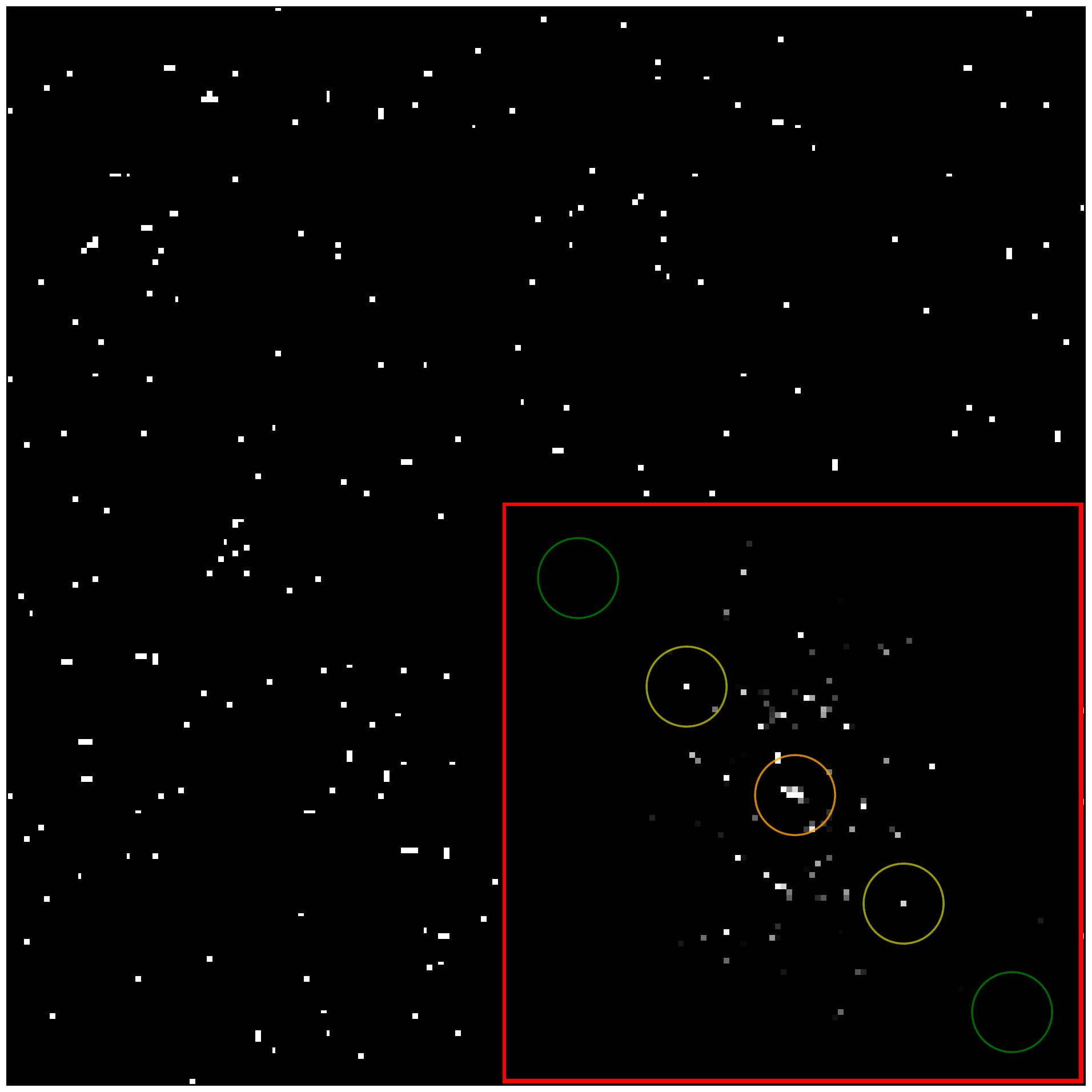}}
    \subfloat[]{\label{subfig:im2}
      \includegraphics[width=0.5\columnwidth]{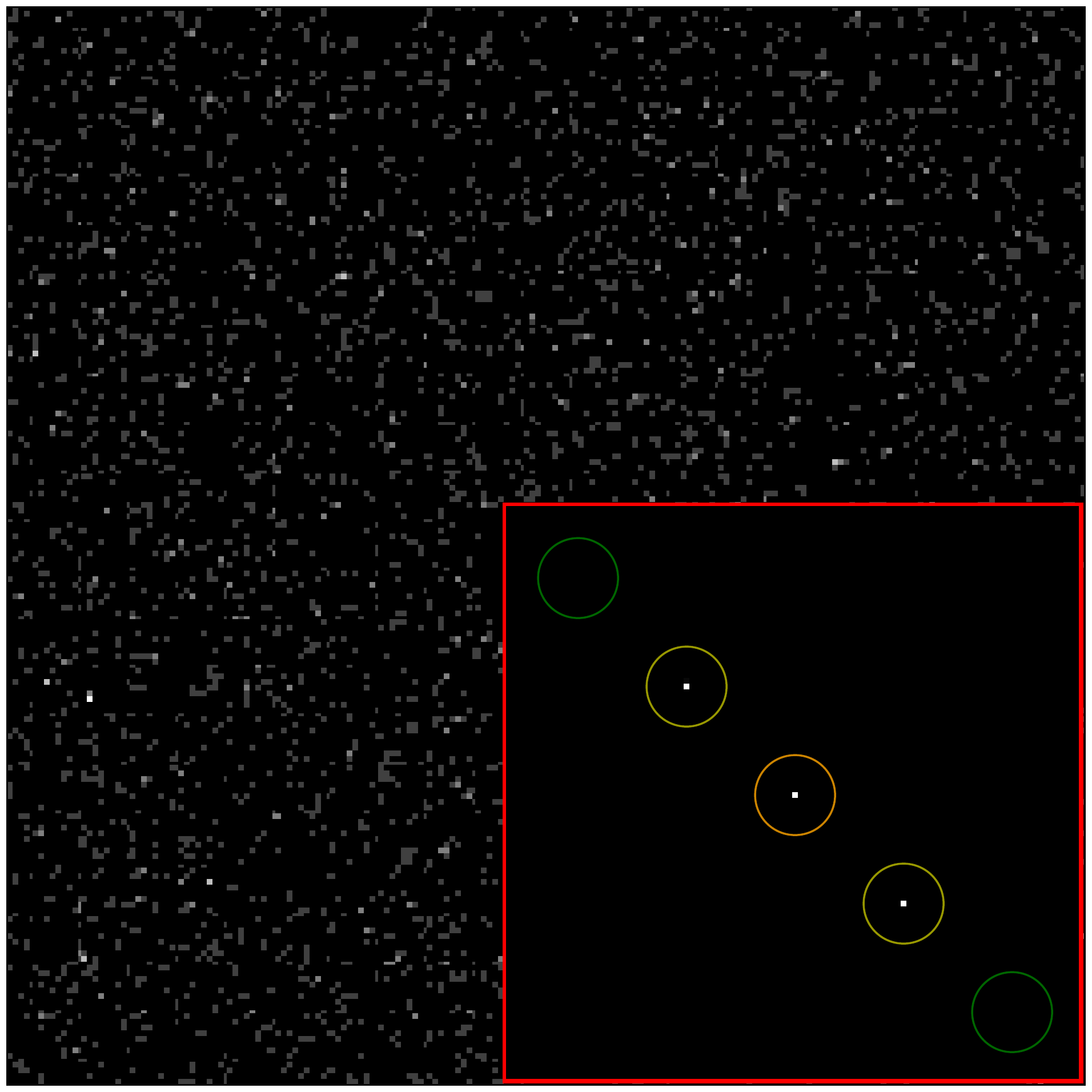} } \\
    \subfloat[]{\label{subfig:im3}
      \includegraphics[width=0.5\columnwidth]{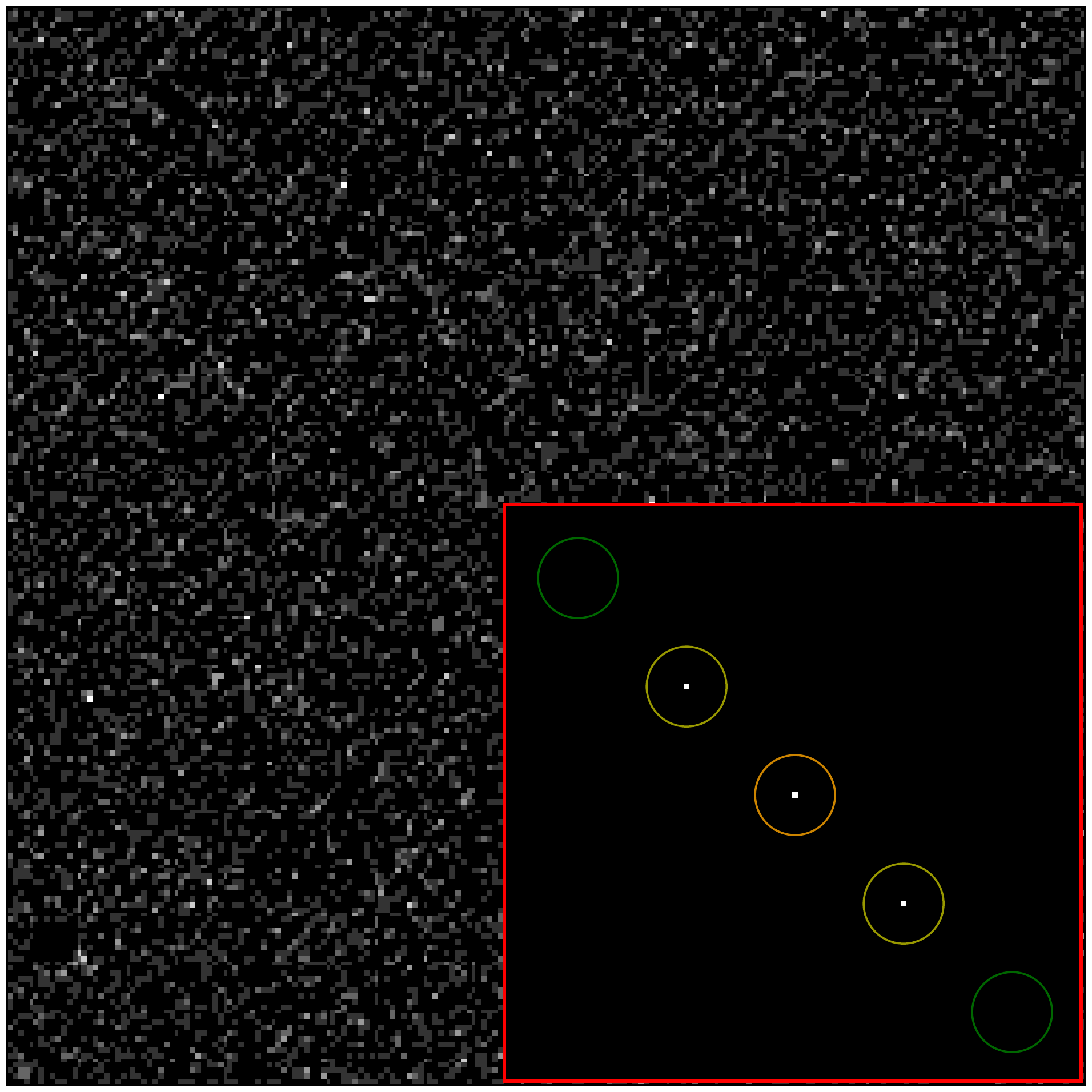}}
    \subfloat[]{\label{subfig:im4}
      \includegraphics[width=0.5\columnwidth]{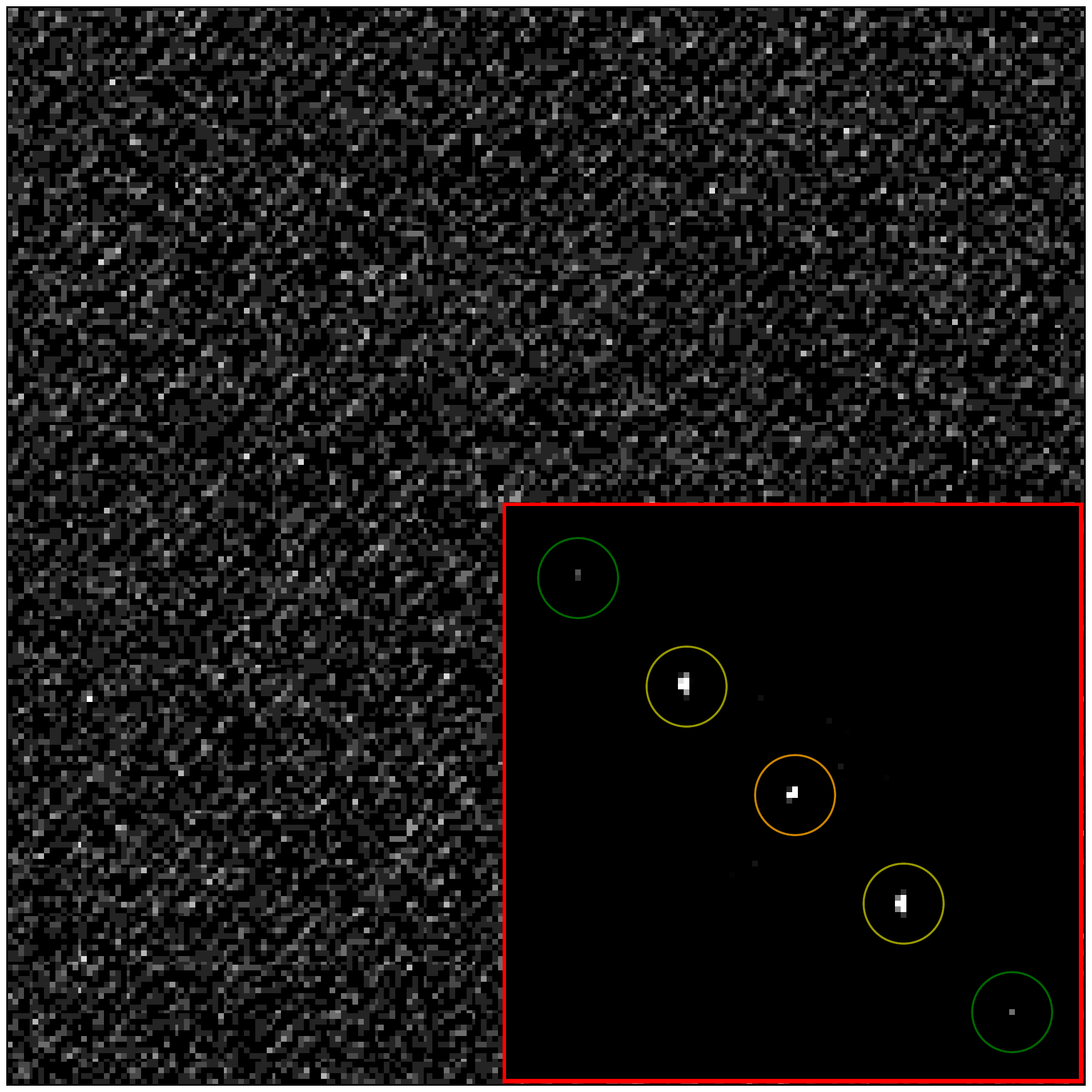} }
    \end{tabular}
  \caption{
    (a-d) Build-up of single electron events resulting in interference fringes after (a) $\unit[0.0025]{\textrm{s}}$, (b) $\unit[0.045]{\textrm{s}}$, (c) $\unit[0.1425]{\textrm{s}}$, and (d) $\unit[0.2875]{\textrm{s}}$. The FFT of each frame is shown in the inset image. Note that frames with $\unit[0.0025]{\textrm{s}}$ exposure were used to reconstruct the phase image shown in Figure \ref{fig:STEMH_Au_C}.
  \label{fig:e_counting} }
\end{figure}
    
    As shown in Figure \ref{subfig:im1}, it is hard to make out interference fringes from a single frame exposed to a beam current of $\unit[0.041]{\mathrm{nA}}$. Increasing the detector's exposure time to acquire a greater number of events results in fringes discernible to the human eye, as in Figures \ref{subfig:im2}-\ref{subfig:im4} and \ref{fig:fringe_vis}. Using a computer, however, we can resolve the fringes in a single frame via a Fourier Transform, and so direct electron detectors have decreased STEM convergent beam electron diffraction (CBED) recording time, therefore decreasing the electron dose seen by the specimen.
    
    We inserted a selected area aperture in an image plane of the diffraction probes in order to reduce noise due to unwanted high-angle scattering. This large aperture only blocks high order diffraction probes ($> 4^{th}$ order) which are assumed to be negligible. The passed probes are then recombined through the post specimen optics and interfere in the image plane on the detector. The phase information of each location on the specimen is extracted by a post process described in the Theory and Reconstruction section below.

    We performed this experiment on TEAM I, an FEI Titan 80-300 operated at 300 KeV in STEM mode with both probe and image aberration correction and a semi-angle of $\unit[30]{\textrm{mrad}}$. A $\unit[50]{\textrm{um}}$ diameter, $\unit[200]{\textrm{nm}}$ pitch sinusoidal phase grating is positioned in the Condenser 3 aperture plane. We imaged a specimen consisting of Au nanoparticles on a thin, amorphous carbon support. The images shown in Figures \ref{subfig:Au10_phi_im} and \ref{subfig:Au9_phi_im} are reconstructed from a 128 x 296 and 115 x 300 2D scan of 1920 x 1792 images, forming two 4D data sets with a field of view of 11.1 nm x 25.8 nm and 8.6 nm x 22.5 nm, respectively.
    
\section{Theory and Reconstruction}

The pre-specimen probe wavefunction is defined to be

\begin{align}
  \psi_i(\mathbf{x}) &= a(\mathbf{x}-\mathbf{x}_{p})
                     &= \sum c_{n} a_{n}\left( \mathbf{x} - \mathbf{x}_{p} - n \mathbf{x}_{0}\right)
\end{align}

\noindent where $\mathbf{x}_p$ is the offset-position of our probe, $a_{n}$ is the phase and intensity distribution of the $n^{th}$ diffraction order, $c_{n}$ is the complex amplitude of the $n^{th}$ diffraction order probe, and $\mathbf{x}_{0}$ is the real-space path separation of any one diffraction order probe from it's nearest neighbor. Note that the grating could, in principle, incorporate holographic designs \cite{harvey_2014} that produce different phase and intensity distributions in each diffraction order, such as vortex beams \cite{mcmorran_2011} or aberration-corrected beams \cite{linck_aberration_2017, shiloh_spherical_2018}. In these experiments, we used a large, straight grating within the aperture, encoding flat phase structure in the probes such that each term in $a(\mathbf{x}-\mathbf{x}_{p})$ describes a sharply-peaked, symmetric function that only differs by a linear phase, or $a_{n} = a_{0}$.

Recall that the probes are scanning through space at the specimen plane, which is why an offset-position of the probe is needed here. We'll use a specimen transfer function $t(\mathbf{x})$ resulting in a post-specimen wavefunction
\begin{equation}
  \psi_f(\mathbf{x}) = a(\mathbf{x}-\mathbf{x}_p) \cdot t(\mathbf{x}),
\end{equation}

\noindent where $t(\mathbf{x})$ is the object transmission function. The far field interference pattern at the detector at probe position $\mathbf{x}_p$ is then

\begin{align} \label{eq:grating_image}
  I_p(\mathbf{k}) &= \left|\Psi_f(\mathbf{k})\right|^2_{p} \nonumber\\
                  &= \left(A_p^{*}(\mathbf{k}) \otimes T^{*}(\mathbf{k})\right) \left( A_p(\mathbf{k}) \otimes T(\mathbf{k})\right),
\end{align}

\noindent where $\otimes$ represents convolution and $\ast$ represents complex conjugate. We use lower-case and capitilized letters to denote real versus reciprocal space variables, respectively.


Now lets make the assumption that there are only three beams, or $c_{\lvert n\rvert > 1}=0$; $n \in [-1,0,1]$, and that only $probe_{+1}$ interacts with the specimen with the other two being reference beams passing through vacuum. Taking the Fourier Transform of \ref{eq:grating_image} results in five sharp peaks, which are visible in insets to Figures \ref{subfig:im1}, \ref{subfig:im2}, \ref{subfig:im3} and \ref{subfig:im4}.


\begin{align} \label{eq:g_final_3_beams_simple}
  \scriptstyle\mathcal{I} \textstyle _{p}(\mathbf{x}) = \scriptstyle\mathcal{I} \textstyle _{-2}\left(\mathbf{x}_{p}, \mathbf{x}\right) + \scriptstyle\mathcal{I} \textstyle _{-1}\left(\mathbf{x}_{p}, \mathbf{x}\right) + \nonumber\\
  \scriptstyle\mathcal{I} \textstyle _{0}\left(\mathbf{x}_{p}, \mathbf{x}\right) + \scriptstyle\mathcal{I} \textstyle _{+1}\left(\mathbf{x}_{p}, \mathbf{x}\right) + \scriptstyle\mathcal{I} \textstyle _{+2}\left(\mathbf{x}_{p}, \mathbf{x}\right)
\end{align}

Equation \ref{eq:g_final_3_beams_simple} is expanded into its full form in the appendix. We can extract the specimen's transfer function by integrating around one of the sharp peaks, along the variable $\mathbf{x}$, which would leave us with the transfer function of the scan position variable $\mathbf{x}_{p}$. We could do this for each peak in $\scriptstyle\mathcal{I} \textstyle _{p}(\mathbf{x})$, which would give us redundant information for peaks that include a signal from more than one of the interferometers that includes the scanning probe interacting with the specimen. For example, if $probe_{+1}$ is the interaction scanning probe, the object transmission function information probed by the interaction scanning probe is encoded in fringes with spacing $k_{0} = \frac{1}{\lvert \mathbf{x}_{0} \rvert}$ due to interference between the $probe_{+1}$ and $probe_{0}$. This period corresponds to the $-1$- and $+1$- order peaks in $\scriptstyle\mathcal{I} \textstyle _{p}(\mathbf{x})$, from which the transmission function can be extracted. 

This information is also encoded in fringes with spacing $k_{0} = \frac{1}{2 \lvert \mathbf{x}_{0} \rvert}$ due to interference between $probe_{+1}$ and $probe_{-1}$, and can therefore be extracted from the $-2$- and $+2$-order peaks in $\scriptstyle\mathcal{I} \textstyle _{p}(\mathbf{x})$. In summary, for a three beam interferometer in which one first order diffraction probe interacts with the specimen, the object transmission function information is stored in both the the first and second orders, respectively, of the Fourier transform of the interference fringe image.

A non-negligible $+2$-order diffraction probe $probe_{+2}$ complicates this picture, and the $-2$ and $+2$ peaks in $\scriptstyle\mathcal{I} \textstyle _{p}(\mathbf{x})$ also contain that information via interference with $probe_{0}$. Because the nanofabricated gratings are designed such that $c_{n>1}$ should be weak, we assume that it is negligible.   

We can also make the assertion that the specimen function in vacuum is just $1$, simplifying equation \ref{eq:g_final_3_beams_simple} even further. Integrating around $\scriptstyle\mathcal{I} \textstyle _{+1}\left(\mathbf{x}_{p}, \mathbf{x}\right)$ in equation \ref{eq:g_final_3_beams_simple}, using $a_{0}\left(\mathbf{x}\right)$ as a kernel, and noting that $A_{0}\left(\mathbf{k}\right)$ is a circular aperture, we arrive at the solution.

\begin{align} \label{eq:g1_int}
    \int_{\Omega \left( + \mathbf{x}_{0} \right)} a_{0}\left(\mathbf{x}\right) &\scriptstyle\mathcal{I} \textstyle  _{+1}\left(\mathbf{x}_{p}, \mathbf{x}\right) d\mathbf{x} \nonumber \\
    &= c^{*}_{0}c_{+1} h\left(\mathbf{x}_{p} \right) \otimes t^{*} \left(\mathbf{x}_{0} + \mathbf{x}_{p}\right)
\end{align}

\noindent where $h\left(\mathbf{x}_{p} \right) = \lvert a_{0}\left(\mathbf{x}_{p} \right) \rvert^{2}$. The full derivation is provided in both the appendix and another submitted manuscript that provides a full treatment of the general theory of STEMH \cite{harvey_STEMH_2018}.

To summarize the numerical object wave reconstruction procedure:

\begin{enumerate}
	\item At each probe position, take the Fourier transform of the interference fringe pattern, resulting in equation \eqref{eq:g_final_3_beams_simple}.
	\item Isolate a small (we used $\unit[< 10 \times 10]{\mathrm{pix}^{2}}$) region around a peak that contains the desired object wave information, $\scriptstyle\mathcal{I} \textstyle _{+1}\left(\mathbf{x}_{p}, \mathbf{x}\right)$.
	\item Define a kernel $a_{0}(\mathbf{x})$ by taking the Fourier transform of a reference image of the interference fringes, i.e. an image when all three probes pass through vacuum, and isolate a small region around the center peak.
	\item Multiply these two peaks and integrate, taking the complex conjugate, equation \eqref{eq:g1_int}.
	\item Repeat for each pattern in the scan, i.e. each $x_{p}$ value.
\end{enumerate}

\subsection{Phase-thickness relation}

The specimen transfer function contains an amplitude and phase, which can be used to calculate the thickness of a specimen. For a non-magnetic specimen, the phase imparted onto an electron wave-front is proportional to the electrostatic potential projected through the bulk of the specimen \cite{yasin_2018}. For amorphous materials, we may consider only the mean inner potential, $V_{i}$. Thus, 

\begin{equation} \label{eq:phi_t}
  \phi = C_{E} V_{i} T\left(\mathbf{x}_{p}\right),
\end{equation}

\noindent where $T\left(\mathbf{x}_{p}\right)$ is the thickness of the specimen for each location in the scan, $\mathbf{x}_{p}$, $C_{E} = \frac{2 \pi}{\lambda} \frac{e}{E} \frac{E_{0} + E}{2 E_{0} + E}$, $\lambda$ is the relativistic wavelength of the electron, $\unit[1.97]{\mathrm{pm}}$ for $E=\unit[300]{\mathrm{keV}}$, where $E$ is the kinetic energy of the electron, $E_{0}$ is the rest energy of the electron, and $e$ is the electron unit charge.

\subsection{Phase uncertainty}

The theory of phase detection uncertainty in electron holography has been worked out in detail by Lichte et al. and de Ruijter et al. \cite{lichte_1987, ruijter_1993}, whose work was experimentally supported by Harscher and Lichte \cite{harcher_1996}. If we only consider the counting statistics for the number of electrons per unit area of the detector at any time (shot noise), the standard deviation for detection of the phase from interference fringes with visibility $\mathcal{V} = \frac{I_{max} - I_{min}}{I_{max} + I_{min}}$ is

\begin{equation} \label{eq:sigma_phi_shot}
    \sigma_{\phi_{th}} = \sqrt{\frac{2}{\mathcal{V}^{2} N}},
\end{equation}

\noindent where $N$ is the number of electrons in the measurement area.

\begin{figure}[h]
    \centering
    \includegraphics[width=0.98\columnwidth]{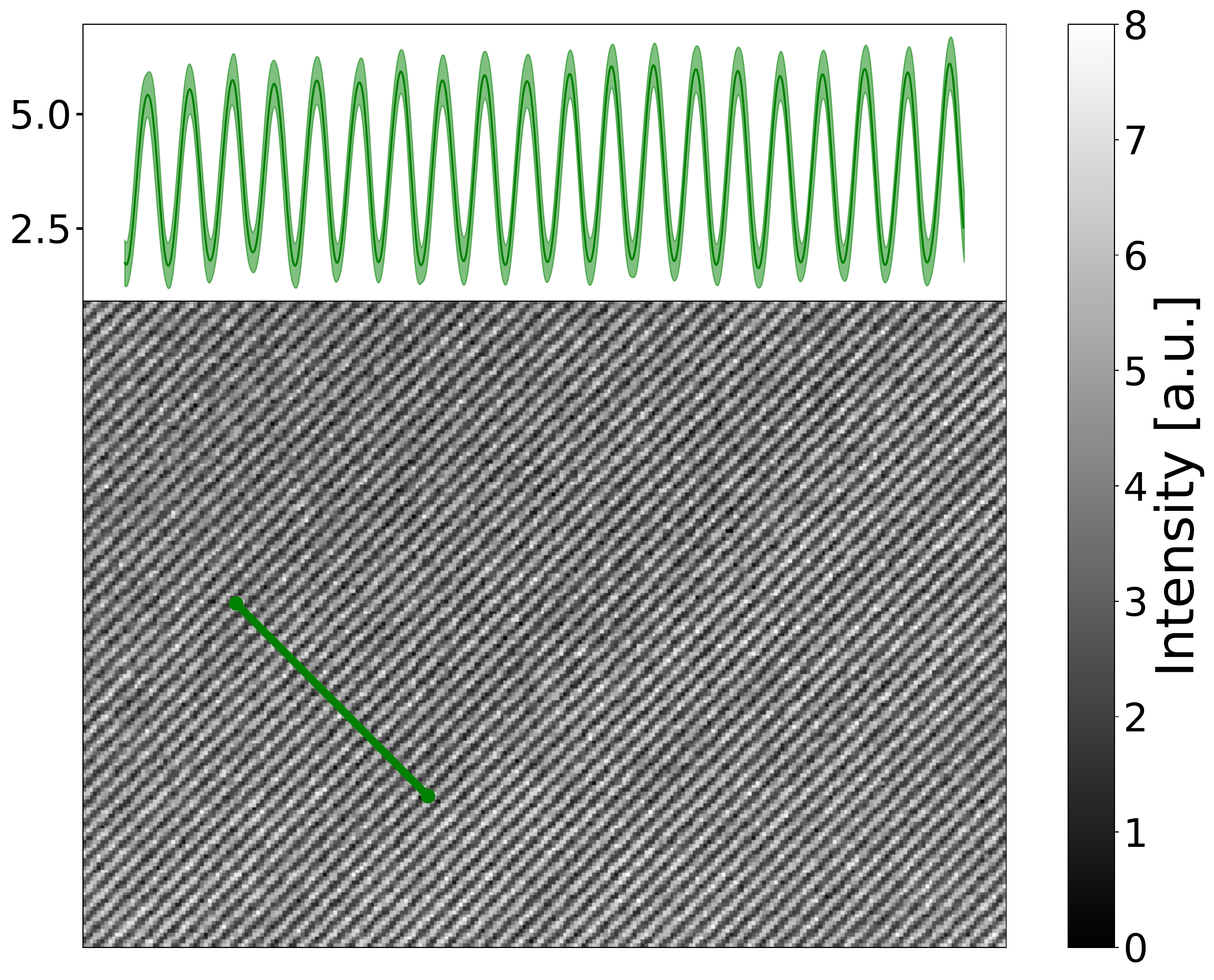}
    \caption{
    Mean interference fringes averaged over the scan with the background subtracted. The inset shows a 1D profile of the sum of fringes along the direction perpendicular to the line trace shown.
    \label{fig:fringe_vis}
    }
\end{figure}

Detectors will also contribute to the phase uncertainty, and their contribution is typically characterized by a detective quantum efficiency, 

\begin{align} \label{eq:DQE}
    DQE = \frac{(SNR)^{2}_{out}(u)}{(SNR)^{2}_{in}(u)},
\end{align}

\noindent where $(SNR)_{out}(u)$ and $(SNR)_{in}(u)$ are signal-to-noise ratios at the output and input of the detector as a function of spatial frequency, $u$ \cite{harcher_1996}. The $DQE$ modifies equation \ref{eq:sigma_phi_shot} to be

\begin{align} \label{eq:sigma_phi}
    \sigma_{\phi_{th}} = \left(DQE\right)^{-\frac{1}{4}}\sqrt{\frac{2}{\mathcal{V}^{2} N}}.
\end{align}

For our experiment, the number of electrons per frame was estimated by summing the intensity values in a frame to be $N \approx 10^{5}$ and we measured our fringe visibility from Figure \ref{fig:fringe_vis} to be $\mathcal{V} = 42.7\% \pm 4.8$. The predicted fringe visibility from an ideal three beam interferometer depends on the phase imparted onto $probe_{+1}$. The fringe spacing at the camera was $\approx 0.38 \times f_{N}$, where $f_{N}$ is the Nyquist frequency. At this spatial frequency, the Gatan K2 Summit camera has a $DQE \approx 0.56$ \cite{ruskin_2013}. Using these values, we plot the numerically calculated $\mathcal{V}$ and $\sigma_{\phi_{th}}$ in the appendix. The mean theoretical uncertainty in phase measurement is $\sigma_{\phi_{th}} < \unit[15]{\mathrm{mrad}}$ when $probe_{+1}$ transmits through a weak phase object.
\clearpage

\onecolumngrid

\begin{figure}[h]
    \centering

    \begin{tabular}{cc}
    \hspace{-8em}
    \adjustbox{valign=b}{\begin{tabular}{cc}
    \subfloat[]{\label{subfig:Au10_H_im}
      \includegraphics[width=0.65\linewidth]{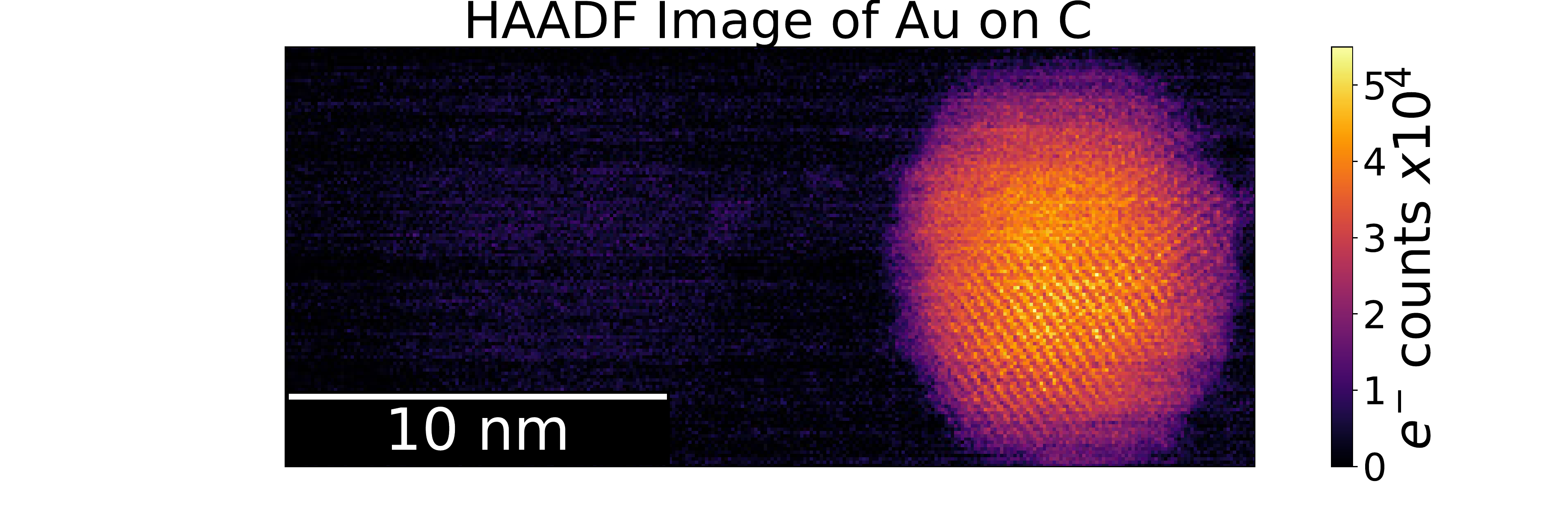} } &
    \hspace{-10em}
    \vspace{-3ex}
    \subfloat[]{\label{subfig:Au10_phi_im}
      \includegraphics[width=0.63\linewidth]{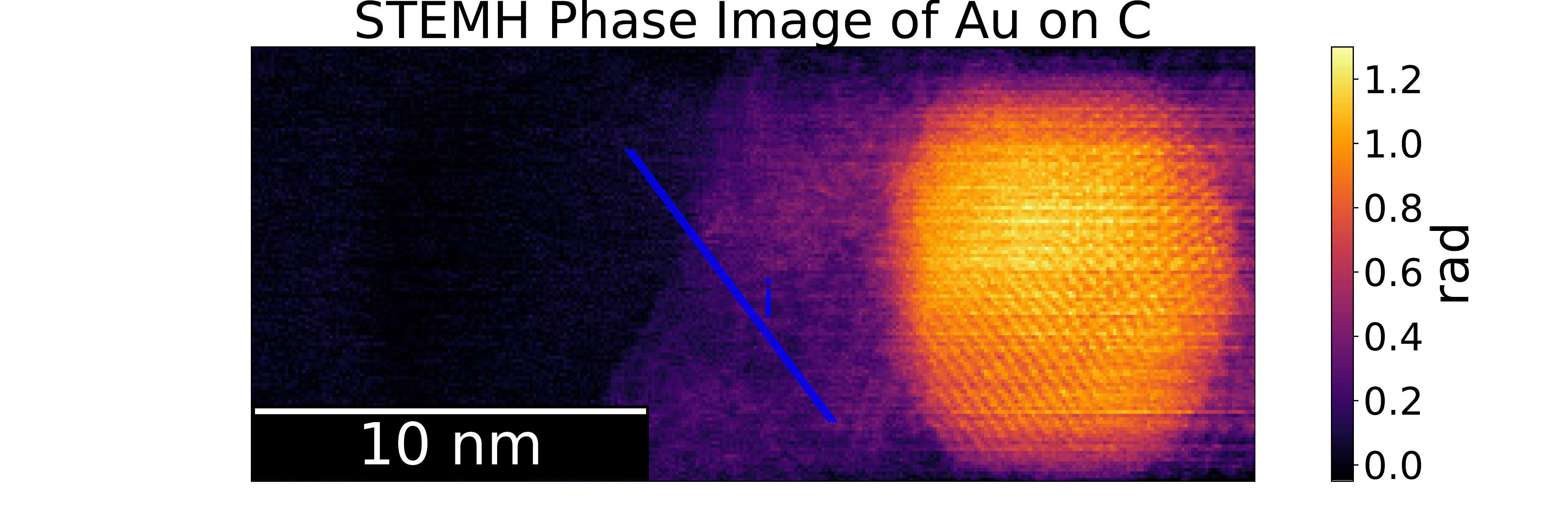}} \\
    \hspace{-1.em}
    \vspace{-1ex}
    \subfloat[]{\label{subfig:Au10_fft_h}
      \includegraphics[width=0.415\linewidth]{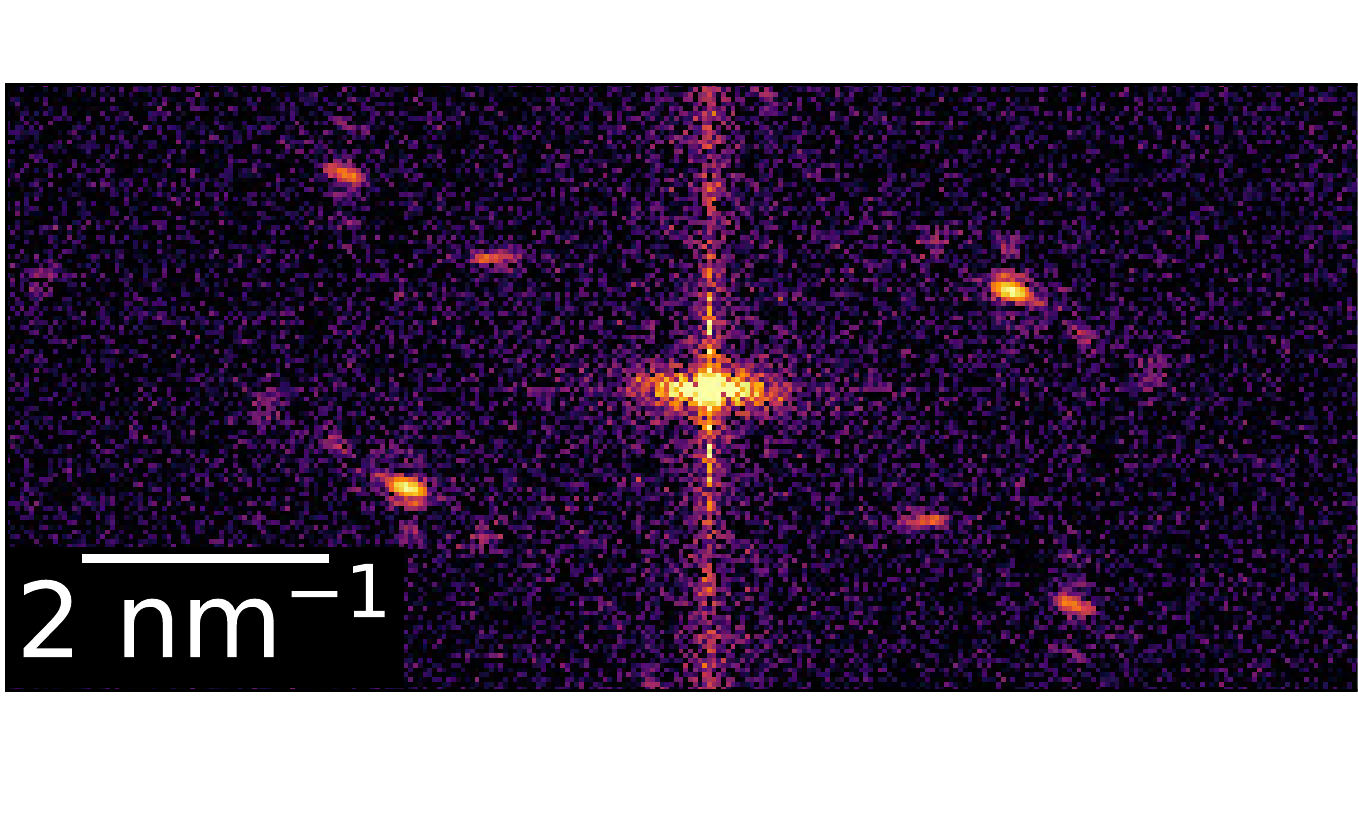} } &
    \hspace{-12.em}
    \vspace{-1ex}
    \subfloat[]{\label{subfig:Au10_fft}
      \includegraphics[width=0.415\linewidth]{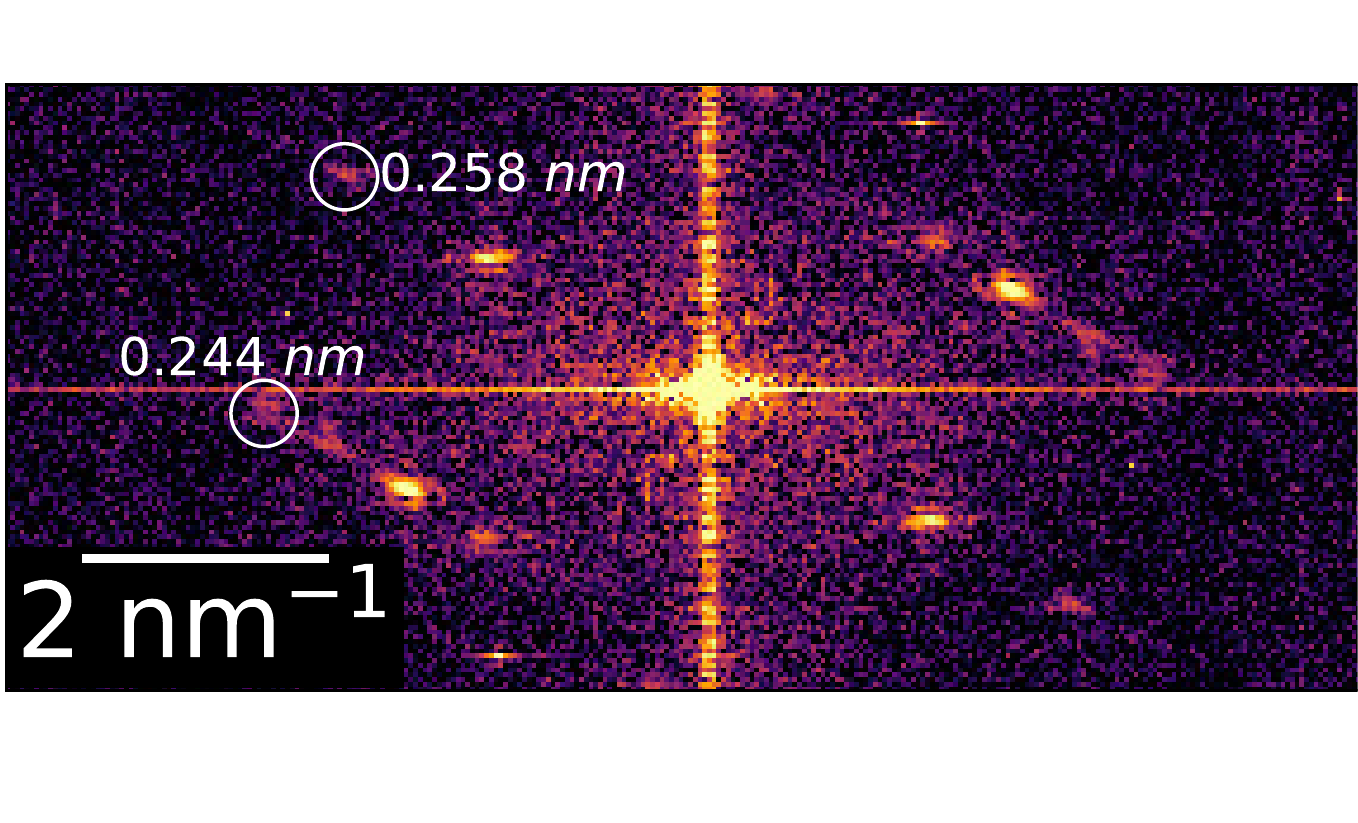}} \\
    \vspace{-1ex}
    \subfloat[]{\label{subfig:Au9_H_im}
      \includegraphics[width=0.65\linewidth]{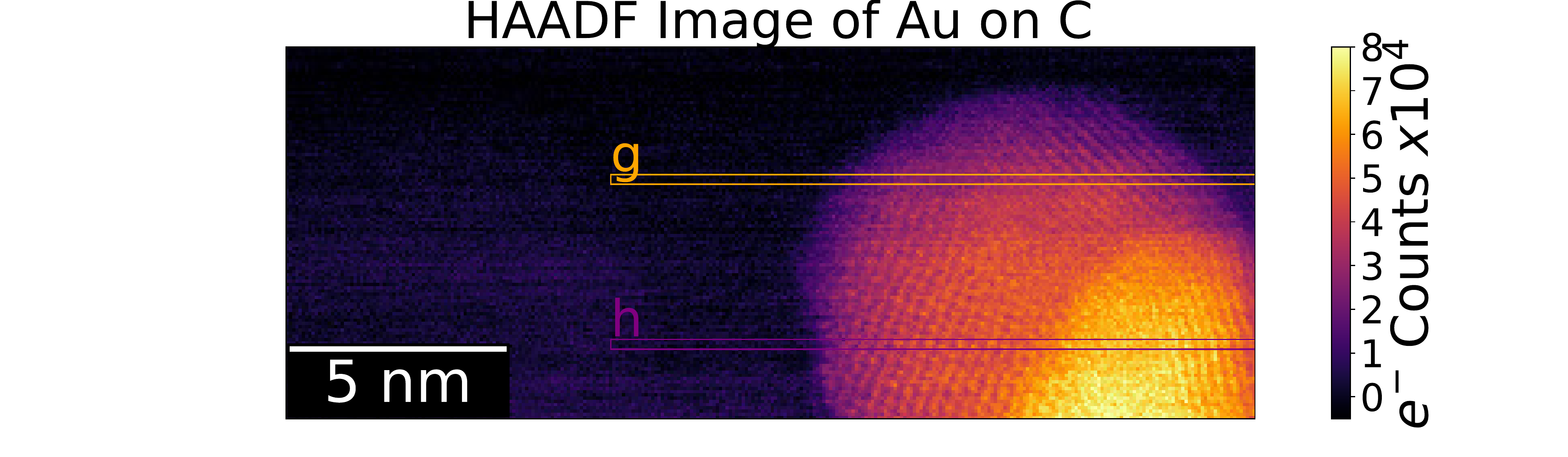}} &
    \hspace{-9.5em}
    \vspace{-1ex}
    \subfloat[]{\label{subfig:Au9_phi_im}
      \includegraphics[width=0.63\linewidth]{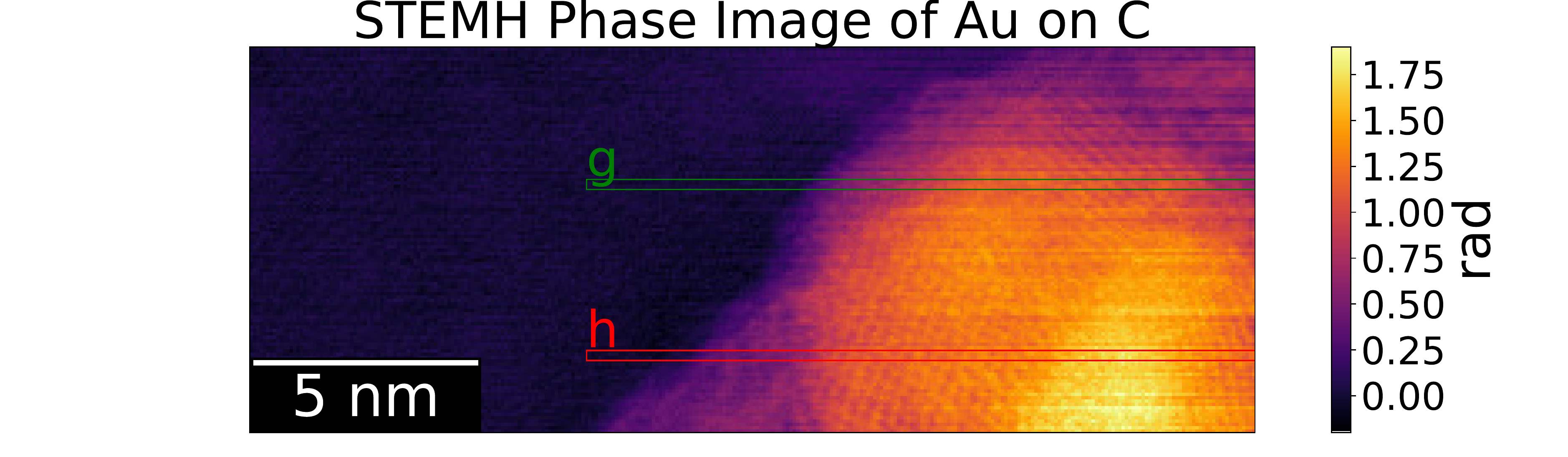}} \\
    \end{tabular}} \\
    \adjustbox{valign=b}{\begin{tabular}{ccc}
    \hspace{-6ex}
    \subfloat[]{\label{subfig:Au9_comp1}
      \includegraphics[width=0.31\linewidth]{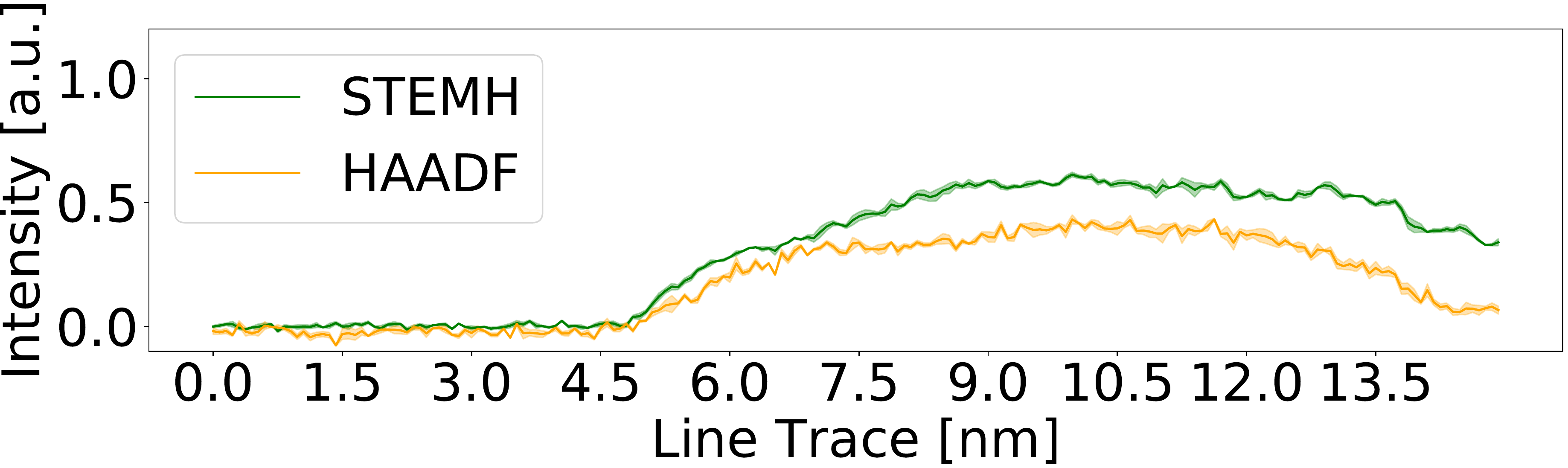}} &
    \hspace{-2ex}
    \subfloat[]{\label{subfig:Au9_comp2}
      \includegraphics[width=0.31\linewidth]{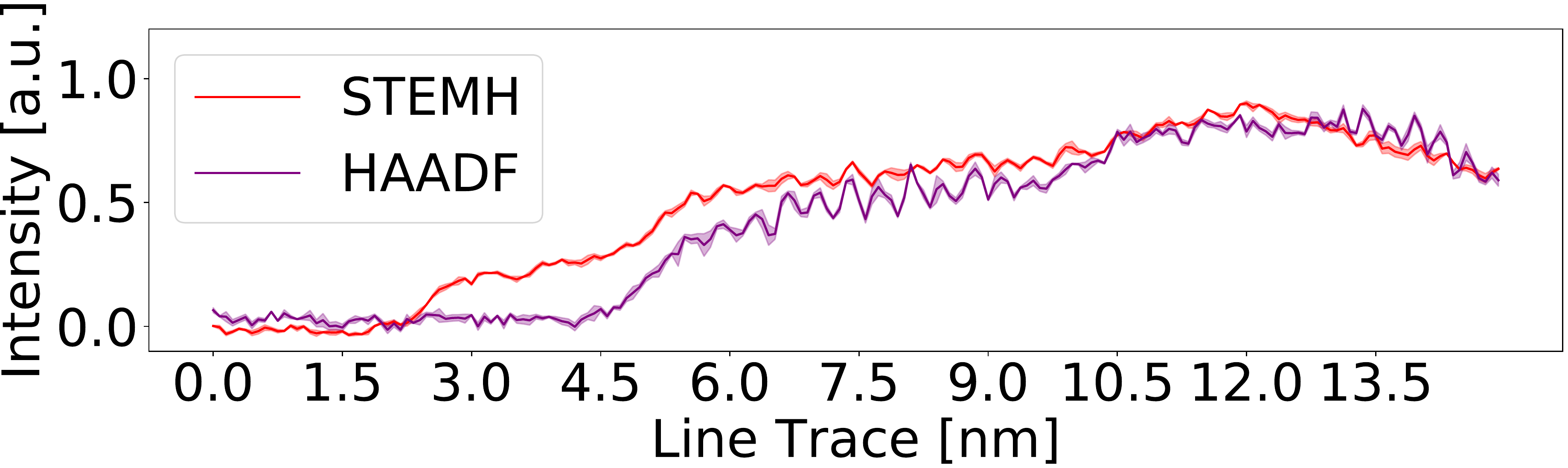} } &
    \hspace{-4.5ex}
    \vspace{-1ex}
    \subfloat[]{\label{subfig:C_phi_t}
      \includegraphics[width=0.35\linewidth]{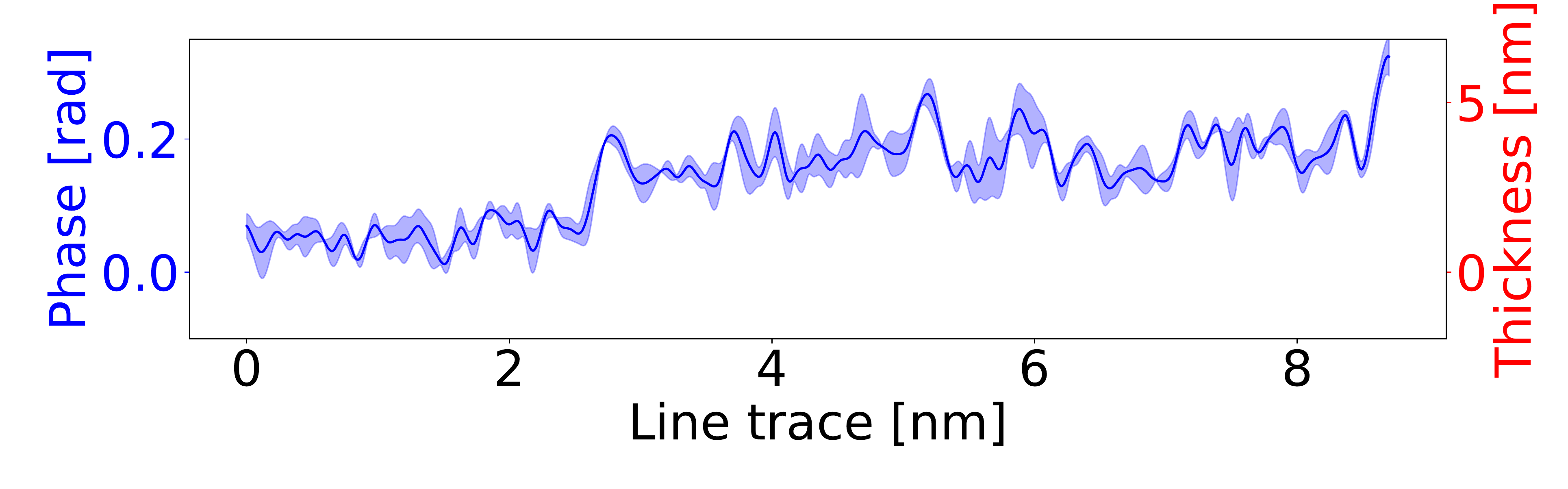}}
    \end{tabular}}

    \end{tabular}
  \caption{
    (a,e) Conventional annular dark field images of two different Au nanoparticles on thin C support. (b,f) Phase reconstruction of the same regions using STEMH. 
    (c-d) Corresponding Fourier Transforms of (a) and (b).
    (g-h) Selected line traces from (e-f), highlighting the low atomic number material contrast seen using STEMH. The profiles are normalized to the maximum value of each image after offsetting to a mean value of zero in vacuum.
    (i) Selected line trace from (b) along just the carbon substrate, from which the thickness is calculated. (g-i) are plots of the mean  along three line traces with the root mean square of the the deviations shaded.
  \label{fig:STEMH_Au_C} }
\end{figure}
\twocolumngrid

\section{Results}

\subsection{STEMH Phase Contrast}

The STEMH phase reconstructions and HAADF images of two randomly oriented Au nanoparticles embedded on a thin amorphous carbon film are shown in Figure \ref{fig:STEMH_Au_C}. Compared to the HAADF image, STEMH allows for a much higher contrast of the thin amorphous carbon. Additionally, the dc-component of the phase is reconstructed using STEMH, resulting in a comparable signal with the HAADF, but with additional amorphous carbon signal barely visible in the HAADF. Figure \ref{subfig:Au10_fft} shows that under the experimental conditions we used, STEMH has $\unit[0.24]{\textrm{nm}}$ resolution of the Au atomic lattice, comparable to the HAADF resolution shown in \ref{subfig:Au10_fft_h}.  Notice how the high frequency information between the two techniques are comparable, whereas the STEMH reconstruction contains much more low frequency information because of the higher contrast on the carbon substrate. Note that these scans were under-sampled in order to achieve a large field of view and decrease both the scan time and file size. The achievable probe size for STEM is sub-angstrom as discussed in the introduction, suggesting that STEMH should be able to achieve even higher resolution than we report.

Figures \ref{subfig:Au9_comp1} - \ref{subfig:Au9_comp2} shows selected line traces along the carbon film and Au nanoparticles for both the STEMH and HAADF signals. For comparison, the signals are normalized to the maximum value of each image after offsetting to a mean value of zero in vacuum. For \ref{subfig:Au9_comp2} the STEMH signal begins to rise earlier than the HAADF due to the amorphous carbon preceding the Au nanoparticles. In \ref{subfig:Au9_comp1}, the two signals rise simultaneously because the nanoparticle hangs off of the edge of the carbon. However, the STEMH signal continues to rise around $\unit[7]{\mathrm{nm}}$, because unlike HAADF STEM the STEMH signal is sensitive to the carbon film lying beneath the Au nanoparticle. This is due to STEMH's phase contrast, resulting in a gap between the two signals after $\unit[7]{\mathrm{nm}}$.

\begin{figure}[h]

\centering

  \includegraphics[width=0.99\columnwidth]{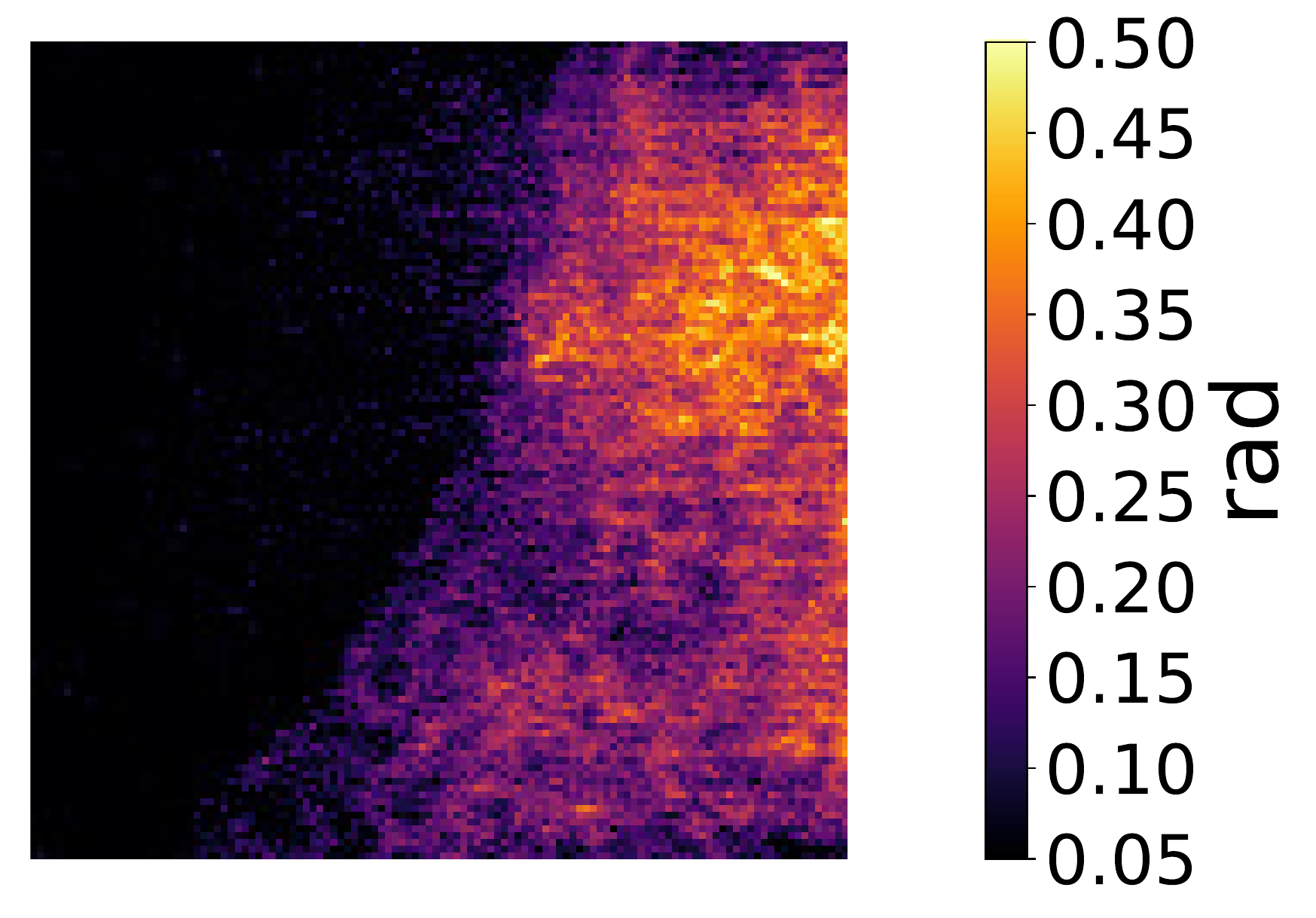} 

  \caption{
    Inset from center of Figure \ref{subfig:Au10_phi_im} in a carbon-only region to enhance contrast.
  \label{fig:C_inset} }

\end{figure}

The line trace in Figure \ref{subfig:Au10_phi_im} follows a path along the amorphous carbon film and is plotted in \ref{subfig:C_phi_t}. The thickness calculated from equation \eqref{eq:phi_t} is shown on the right vertical axis. Interestingly, the carbon in Figures \ref{subfig:Au10_phi_im} and \ref{subfig:Au9_phi_im}, isolated in Figure \ref{fig:C_inset}, shows a string-like topography, which seems to be consistent with a thick-bonding model detailed by Ricolleau, et al. in 2013 \cite{ricolleau_2013}. 

We previously assumed that $probe_{+2}$ was weak, or $c_{+2} < \frac{c_{+1}}{10}$. This results in a phase signal $<\frac{1}{10}$ the $probe_{+1}$ signal. Such a weak signal is present in Figures \ref{subfig:Au10_phi_im} and \ref{subfig:Au9_phi_im} in the form of a `shadow' image of the nanoparticle in vacuum, although it is weak enough to be barely identifiable above the noise from the primary signal on carbon.

As shown in the previous section, however, this signal is on the same order or larger than our theoretical uncertainty, which is confirmed by our measurement of uncertainty in the phase. Here, we measured the deviation from the mean for both a single line and an area of 50 lines within the vacuum region of Figure \ref{subfig:Au10_phi_im}. We found that for a single line, $\sigma_{\phi_{exp}} = \unit[30]{\mathrm{mrad}}$ and $\sigma_{\phi_{exp}} = \unit[35]{\mathrm{mrad}}$ for an area of 50 lines. This difference can be attributed to scan noise. The increase in noise between theory and experiment is consistent with contributions due to higher order probes, and so future consideration should be taken when designing gratings so as to optimize the output SNR. Alternatively, a smaller selected area aperture could be used to block higher orders.

\section{Conclusion}

    In this article, we demonstrated sub-nanometer resolution electron phase imaging using STEMH, a multiple-arm, path-separated interferometer with a phase imparted onto one or more paths. We measured a fringe visibility of $\mathcal{V} = 42.7\%$ experimental uncertainty in phase measurement to be $\sigma_{\phi_{exp}} \approx \unit[0.03]{\mathrm{rad}}$. We then provided two $\unit[0.24]{\textrm{nm}}$ resolution phase-contrast images of Au nanoparticles on a thin carbon substrate, with conventional HAADF images for comparison. 
    
    STEMH provides quantitative phase contrast, including the dc-component, which we utilized to analyze the thickness of the carbon support. Recall that we used a straight grating in this experiment to prepare sharply-peaked, symmetric probes at the sample plane. Note, however, that more complicated diffraction grating designs can be used to holographically vary the complex amplitude $c_{n}$ of the diffraction orders \cite{grillo_2014, harvey_2014, shiloh_2014, ophus_efficient_2016}, potentially enabling more complicated electron-specimen interactions with signals extractable via STEMH. Future additions of faster readout detectors and different grating designs would further reduce the electron dose, potentially allowing STEMH to image beam-sensitive, bio-molecular materials at atomic resolution.


\section*{Acknowledgements}
\begin{acknowledgments}
This work was supported by both the U.S. Department of Energy, Office of Science, Basic Energy Sciences, under Award DE-SC0010466 and by the National Science Foundation Graduate Research Fellowship Program under Grant No. 1309047, with partial support by the National Science Foundation under award 2004Y1. 
\end{acknowledgments}

\bibliographystyle{apsrev4-1}
\bibliography{180727_STEMH_NCEM}{}

\begin{widetext}
\section{Appendix}
\subsection{Full form of equation \ref{eq:g_final_3_beams_simple} }

\begin{align} \label{eq:g_3_beams}
    \scriptstyle\mathcal{I} \textstyle _p(\mathbf{x}) = &\sum_{n=-1}^{+1} \lvert c_{n} t\left(n \mathbf{x}_{0} + \mathbf{x}_{p}\right) \rvert^{2} +  c^{*}_{-1}c_{0} \left[ a^{*}_{-1}\left( \mathbf{x} + 1 \mathbf{x}_{0} \right) t^{*} \left( \mathbf{x} + \mathbf{x}_{p}\right) \right] \otimes \left[ a_{0}\left( \mathbf{x} + 1 \mathbf{x}_{0} \right) t \left( \mathbf{x} + 1\mathbf{x}_{0} - \mathbf{x}_{p}\right) \right] \nonumber \\
                    + &c^{*}_{0}c_{-1} \left[ a^{*}_{0}\left( \mathbf{x} - 1 \mathbf{x}_{0} \right) t^{*} \left( \mathbf{x} - 1\mathbf{x}_{0} + \mathbf{x}_{p}\right) \right] \otimes \left[ a_{-1}\left( \mathbf{x} - 1 \mathbf{x}_{0} \right) t \left( \mathbf{x} - \mathbf{x}_{p}\right) \right] \nonumber \\
                    + &c^{*}_{-1}c_{+1} \left[ a^{*}_{-1}\left( \mathbf{x} + 2 \mathbf{x}_{0} \right) t^{*} \left( \mathbf{x} + 1\mathbf{x}_{0} + \mathbf{x}_{p}\right) \right] \otimes \left[ a_{+1}\left( \mathbf{x} + 2 \mathbf{x}_{0} \right) t \left( \mathbf{x} + 1\mathbf{x}_{0} - \mathbf{x}_{p}\right) \right] \nonumber \\
                    + &c^{*}_{+1}c_{-1} \left[ a^{*}_{+1}\left( \mathbf{x} - 2 \mathbf{x}_{0} \right) t^{*} \left( \mathbf{x} - 1\mathbf{x}_{0} + \mathbf{x}_{p}\right) \right] \otimes \left[ a_{-1}\left( \mathbf{x} - 2 \mathbf{x}_{0} \right) t \left( \mathbf{x} - 1\mathbf{x}_{0} - \mathbf{x}_{p}\right) \right] \nonumber \\
                    + &c^{*}_{0}c_{+1} \left[ a^{*}_{0}\left( \mathbf{x} + 1 \mathbf{x}_{0} \right) t^{*} \left( \mathbf{x} + 1\mathbf{x}_{0} + \mathbf{x}_{p}\right) \right] \otimes \left[ a_{+1}\left( \mathbf{x} + 1 \mathbf{x}_{0} \right) t \left( \mathbf{x} - \mathbf{x}_{p}\right) \right] \nonumber \\
                    + &c^{*}_{+1}c_{0} \left[ a^{*}_{+1}\left( \mathbf{x} - 1 \mathbf{x}_{0} \right) t^{*} \left( \mathbf{x} + \mathbf{x}_{p}\right) \right] \otimes \left[ a_{0}\left( \mathbf{x} - 1 \mathbf{x}_{0} \right) t \left( \mathbf{x} - 1\mathbf{x}_{0} - \mathbf{x}_{p}\right) \right]
\end{align}

Collecting the $\ell^{th}$ peak terms $a\left(\mathbf{x + \ell\mathbf{x}_{0}}\right)$, we can write this in the simpler form seen in equation \ref{eq:g_final_3_beams_simple}.

\subsection{Derivation of transfer function reconstruction}
Let's integrate out the $\mathbf{x}$ variable around the $+1$ order peak in $_{p}(\mathbf{x})$, using $a_{0}$ as a kernel.

\begin{align} \label{eq:g1_int}
    \int_{\Omega \left( + \mathbf{x}_{0} \right)} a_{0}\left(\mathbf{x}\right) \scriptstyle\mathcal{I} \textstyle _{+1}\left(\mathbf{x}_{p}, \mathbf{x}\right) d\mathbf{x} = & \int c^{*}_{0}c_{+1} a_{0}\left(\mathbf{x}\right) \left[ a^{*}_{0}\left( \mathbf{x}\right) t^{*} \left( \mathbf{x} + 1\mathbf{x}_{0} + \mathbf{x}_{p}\right) \right] \otimes a_{+1}\left( \mathbf{x} \right) d\mathbf{x} \nonumber\\
    Using\ the\ commutivity\ of\ convolutions: \nonumber\\
             = &c^{*}_{0}c_{+1}\int \int a_{0}\left(\mathbf{x} \right)a_{+1}\left( \mathbf{x} - \mathbf{x}' \right) \left[ a^{*}_{0}\left( \mathbf{x}'\right) t^{*} \left( \mathbf{x}' + 1\mathbf{x}_{0} + \mathbf{x}_{p}\right) \right] d\mathbf{x} d\mathbf{x}'
\end{align}

Because $a_{n}\left(\mathbf{x}\right)$ is a symmetric function, $\int a_{0}\left(\mathbf{x} \right)a_{+1}\left( \mathbf{x} - \mathbf{x}' \right) d\mathbf{x} = a^{*}_{0}\left(\mathbf{x} \right)' \star a_{+1}\left( \mathbf{x}' \right) = a_{0}\left(-\mathbf{x}' \right) \otimes a_{+1}\left( \mathbf{x}' \right) = a_{0}\left(\mathbf{x}' \right) \otimes a_{+1}\left( \mathbf{x}' \right)$. We can simplify this further using the convolution theorem, and noting that the circular aperture $A_{0}\left(\mathbf{k}\right)$ is a top hat function:

\begin{equation} \label{eq:aperture_top_hat}
  A_{m}\left( \mathbf{k}\right) = A_{0}\left( \mathbf{k}\right) = \begin{cases}
                      \frac{1}{\pi K^{2}} \quad \, \lvert \mathbf{k} \rvert \leq K \\
                      0 \quad \, \lvert \mathbf{k} \rvert > K \\
                  \end{cases}
\end{equation}

\begin{align} \label{eq:g1_int_conv}
   \int_{\Omega \left( + \mathbf{x}_{0} \right)} a_{0}\left(\mathbf{x}\right) \scriptstyle\mathcal{I} \textstyle _{+1}\left(\mathbf{x}_{p}, \mathbf{x}\right) d\mathbf{x} = &c^{*}_{0}c_{+1}\int \int e^{-2 \pi i \mathbf{k} \cdot \mathbf{x}'} A_{0}\left(\mathbf{k} \right) A_{+1}\left( \mathbf{k} \right) \left[ a^{*}_{0}\left( \mathbf{x}'\right) t^{*} \left( \mathbf{x}' + 1\mathbf{x}_{0} + \mathbf{x}_{p}\right) \right] d\mathbf{k} d\mathbf{x}' \nonumber\\
            = &c^{*}_{0}c_{+1}\int \int e^{-2 \pi i \mathbf{k} \cdot \mathbf{x}'} \lvert A_{0}\left(\mathbf{k} \right)\rvert^{2} \left[ a^{*}_{0}\left( \mathbf{x}'\right) t^{*} \left( \mathbf{x}' + 1\mathbf{x}_{0} + \mathbf{x}_{p}\right) \right] d\mathbf{k} d\mathbf{x}' \nonumber\\
            = &c^{*}_{0}c_{+1}\int \int e^{-2 \pi i \mathbf{k} \cdot \mathbf{x}'} A_{0}\left(\mathbf{k} \right) \left[ a^{*}_{0}\left( \mathbf{x}'\right) t^{*} \left( \mathbf{x}' + 1\mathbf{x}_{0} + \mathbf{x}_{p}\right) \right] d\mathbf{k} d\mathbf{x}' \nonumber\\
            = &c^{*}_{0}c_{+1}\int a_{0}\left(\mathbf{x}' \right) \left[ a^{*}_{0}\left( \mathbf{x}'\right) t^{*} \left( \mathbf{x}' + 1\mathbf{x}_{0} + \mathbf{x}_{p}\right) \right]  d\mathbf{x}' \nonumber\\
            = &c^{*}_{0}c_{+1}\int \lvert a_{0}\left(\mathbf{x}' \right) \rvert^{2} t^{*} \left( \mathbf{x}' + 1\mathbf{x}_{0} + \mathbf{x}_{p}\right) d\mathbf{x}' \nonumber\\
            = &c^{*}_{0}c_{+1} \left( \lvert a_{0}\left(\mathbf{x}_{p} \right) \rvert^{2} \right)^{*} \star t^{*} \left( \mathbf{x}_{0} + \mathbf{x}_{p}\right) \nonumber\\
            = &c^{*}_{0}c_{+1} \lvert a_{0}\left(-\mathbf{x}_{p} \right) \rvert^{2} \otimes t^{*} \left(\mathbf{x}_{0} + \mathbf{x}_{p}\right) \nonumber\\
            = &c^{*}_{0}c_{+1} h\left(\mathbf{x}_{p} \right) \otimes t^{*} \left(\mathbf{x}_{0} + \mathbf{x}_{p}\right)
\end{align}

Since $a_{0}$ is symmetric, $h\left(\mathbf{x}_{p} \right) = \lvert a_{0}\left(\mathbf{x}_{p} \right) \rvert^{2}$.

\subsection{Numerical calculation of $\sigma_{\phi_{th}}$}

For an ideal three beam interferometer, the three probes are of equal amplitude ($c_{n} \approx \frac{1}{\sqrt{3}}$). In the following calculation, we simulated a phase grating with the following transmission function:

\begin{equation} \label{grating}
G\left(\mathbf{k}\right) = \exp{\left( \ \Delta \phi \  i \frac{\left(1 + cos\left(\frac{2 \pi}{d} \mathbf{k}\right)\right)}{2} \right) } \times A_{0}(\mathbf{k}),
\end{equation}

\noindent where $\Delta \phi$ is the phase depth, a complex coefficient that determines the diffraction grating efficiency and wavefunction amplitude loss, while d is the grating pitch. For the simulation, we used $\Delta \phi = 2.869$, which corresponds to diffraction probe amplitudes of $c_{n} = 0.299$, for $n \in [-1, 0, 1]$. The grating pitch was $d = \unit[160]{\mathrm{nm}}$ and the diameter was $\unit[50]{\mathrm{\mu m}}$. We then calculated the probe wavefunction and applied a phase to $probe_{+1}$. We calculated the fringe visibility from equation \eqref{V_fft}, which utilizes the fast Fourier transform of the fringe pattern. There are two fringe spacings, and so equation \eqref{V_fft} calculates the fringe visibility of the $m^{th}$ FFT peak, corresponding to the $probe_{+1}$/$probe_{-1}$ interferometer ($m=2$) and the $probe_{+1}$/$probe_{0}$ interferometer ($m=1$).

\begin{equation} \label{V_fft}
\mathcal{V} = \frac{\scriptstyle\mathcal{I} \textstyle _{m} + \scriptstyle\mathcal{I} \textstyle _{m}}{\scriptstyle\mathcal{I} \textstyle _{0}} = \frac{ \Delta I}{2 \langle I \rangle }
\end{equation}

\begin{figure}[ht]
    \centering
    \begin{tabular}{cc}

    \subfloat[]{\label{subfig:V_vs_phi}
      \includegraphics[width=0.5\linewidth]{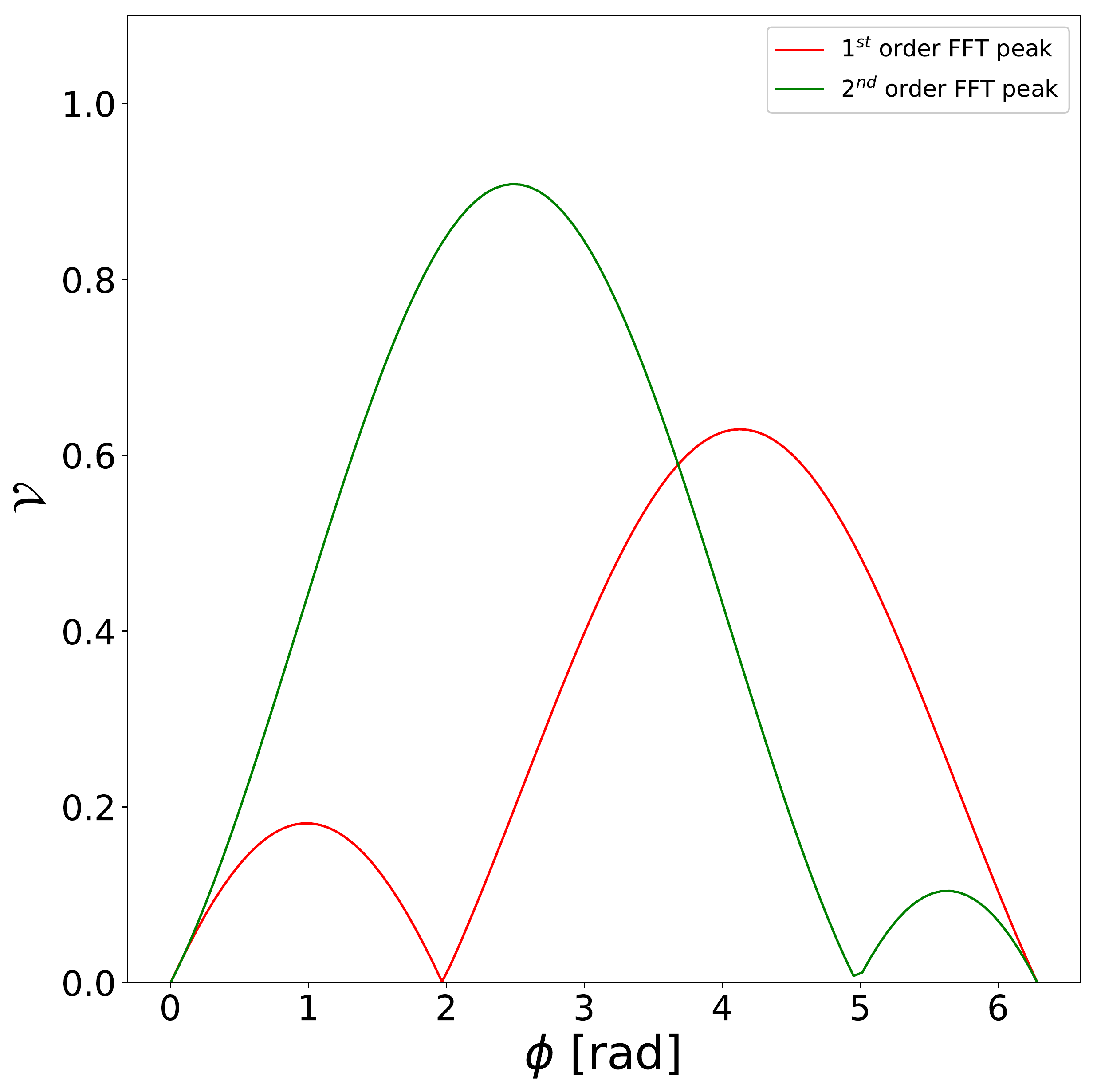}}&
    \subfloat[]{\label{subfig:sig_phi}
      \includegraphics[width=0.5\linewidth]{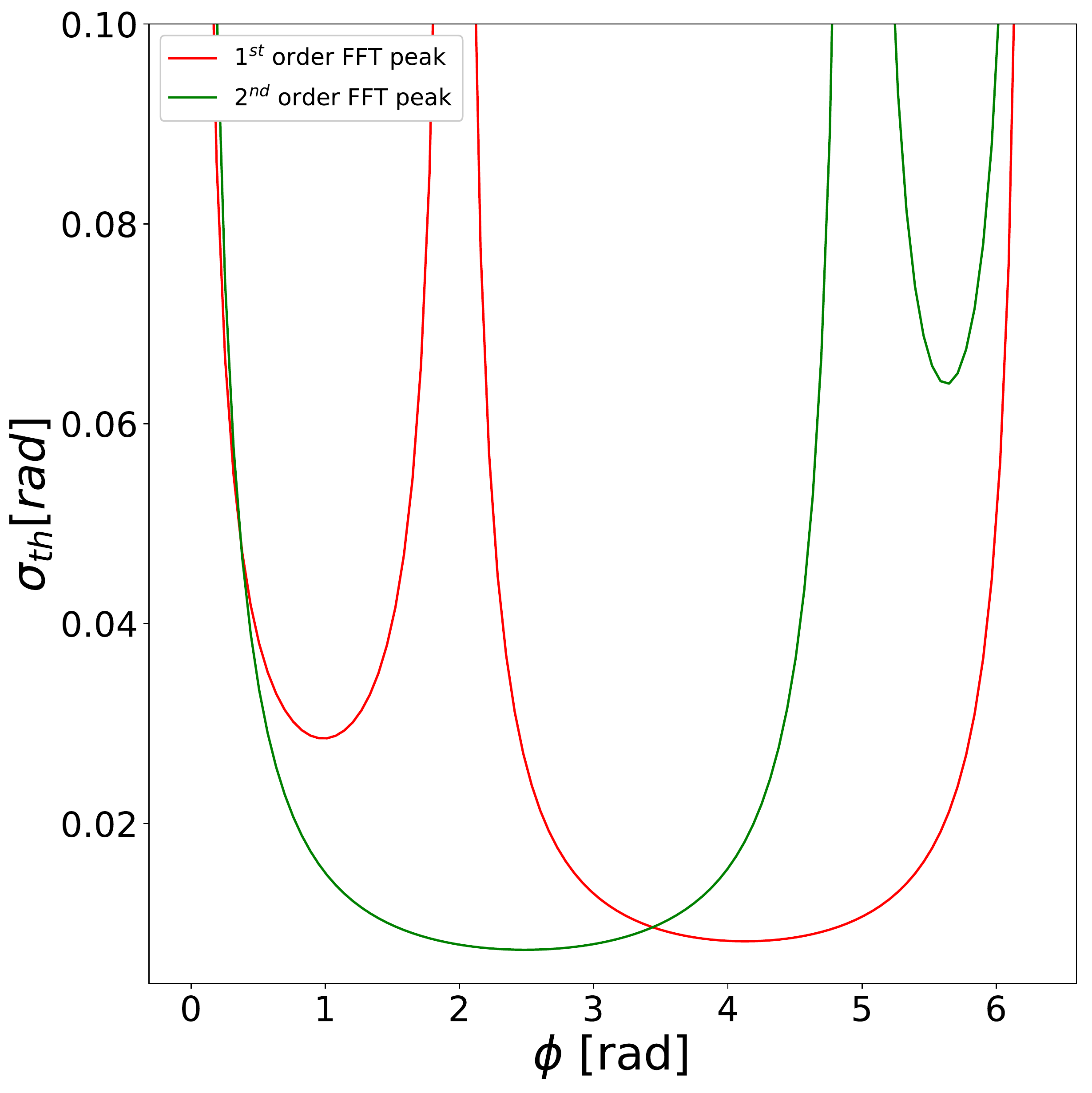} }
    \end{tabular}
  \vspace{-1em}
  \caption{
    (a) The fringe visibility and (b) root-mean-squared uncertainty as a function of phase imparted onto $probe_{+1}$ for both $m=1$ (red) and $m=2$ (green).
  \label{fig:S1} }
\end{figure}

As shown in Figure \ref{subfig:V_vs_phi}, the fringe visibility $\mathcal{V}$ varies between $\unit[0]{\%}$ and $\unit[91]{\%}$. This of course means that for a pure phase grating, the phase uncertainty diverges at $phi = \ell 2 \pi$, where $\ell$ is an integer value. Realistically, these gratings are not ideal phase gratings, and so the visibility is nonzero in vacuum. The corresponding phase uncertainty is shown in Figure \ref{subfig:sig_phi}. Since the phase information is measured in both interferometers, STEMH can utilize both signals to decrease the phase uncertainty over a range of phase values.
\end{widetext}
\end{document}